\documentclass[letter,superscriptaddress,twocolumn,prd]{revtex4-1}
\usepackage{graphicx}
\usepackage{bm}
\usepackage{color}
\usepackage{amsmath}
\usepackage{amssymb}
\usepackage{hyperref}

\newcommand{\hMpc}{\ h^{-1}\text{Mpc}}
\newcommand{\hMpcc}{\ h^{-3}\text{Mpc}^3}
\newcommand{\ihMpc}{\ h\text{Mpc}^{-1}}

\newcommand{\hMs}{\ h^{-1} M_\odot}

\newcommand{\tim}[1]{\times 10^{#1}}

\newcommand{\derd}{\,\mathrm{d}} 
\newcommand{\ddir}{\delta^\text{(D)}}

\newcommand{\be}{\begin{equation}}
\newcommand{\ee}{\end{equation}}
\newcommand{\la}{\left\langle}
\newcommand{\ra}{\right\rangle}

\renewcommand{\vec}{\bm}

\def\mnras{{Mon.\ Not.\ Roy.\ Astron.\ Soc.\ }}

\def\prd{{Phys.\ Rev.\ }}

\def\apj{{Astrophys.\ J.\ }}

\def\apjl{{Astrophys.\ J.\ Let.\ }}

\def\jcap{{JCAP\ }}

\def\lsim{~\rlap{$<$}{\lower 1.0ex\hbox{$\sim$}}}
\def\bsim{~\rlap{$>$}{\lower 1.0ex\hbox{$\sim$}}}

\def\la{\langle}
\def\ra{\rangle}

\def\zvh{\mathrm{\hat{\bf{z}}}}

\def\apjl{Astrophys.\ J.\ Lett.}
\def\mnras{Mon.\ Not.\ R.\ Astron.\ Soc.}

\def\apj{Astrophys.\ J.}

\def\prd{Phys.\ Rev.\ D}

\def\jcap{{JCAP\ }}

\definecolor{orange}{rgb}{1,0.4,0}

\begin{document}

\author{Tobias Baldauf}
\email{baldauf@ias.edu}
\affiliation{School of Natural Sciences, Institute for Advanced Study, Einstein Drive, Princeton, NJ 08540, USA}
\author{Vincent Desjacques}
\affiliation{D\'epartement de Physique Th\'eorique and Center for Astroparticle Physics,
Universit\'e de Gen\`eve, 24 quai Ernest Ansermet, CH-1221 Gen\`eve 4, Switzerland}
\author{Uro\v{s} Seljak}
\affiliation{Department of Physics, University of California Berkeley, CA 94720, USA}
\affiliation{Lawrence Berkeley National Laboratory, Physics Department, Berkeley, CA 94720, USA}
\title{Velocity bias in the distribution of dark matter halos}
\begin{abstract}
The standard formalism for the co-evolution of halos and dark matter predicts that any
initial halo velocity bias rapidly decays to zero.
We argue that, when the purpose is to compute statistics like power spectra etc., the 
coupling in the momentum conservation equation for the biased tracers must be modified.
Our new formulation predicts the constancy in time of any {\it statistical} halo  
velocity bias present in the initial conditions, in agreement with peak theory. 
We test this prediction by studying the evolution of a conserved halo population in  
N-body simulations. 
We establish that the initial simulated halo density and velocity statistics show distinct 
features of the peak model and, thus, deviate from the simple local Lagrangian bias.
We demonstrate, for the first time, that the time evolution of their velocity is in 
tension with the rapid decay expected in the standard approach.
\end{abstract}
\maketitle

\section{Introduction}

The three dimensional late time matter distribution in the Universe has the potential 
to place stringent constraints on cosmological parameters and fundamental physics. 
The interpretation of the observed galaxy distribution is, however, hampered by the fact, 
that galaxies and their host halos are imperfect tracers of the matter distribution.
The local bias model has been successful in explaining the varying clustering amplitude 
observed for different tracers of the large scale structure \cite{Tegmark:2006co}.  
With data from current and upcoming surveys, one would like to push the maximum wavenumber 
in the analysis into the weakly non-linear regime, requiring consistent bias descriptions 
that go beyond the simple local bias model. In this regard, it was recently demonstrated 
that the local bias model is not even consistent with gravitational evolution. Namely, 
tracers which are initially locally biased will always develop nonlocal bias contributions. 
\cite{Chan:2012gr,Baldauf:2012ev}.

In this letter, we focus on a different extension of the local bias model known as the peak 
approach \cite{Bardeen:1985th}, which has its roots in the initial distribution of the 
regions that will eventually form a dark matter halo. 
The peak model predicts the existence of a linear, {\it statistical} halo velocity bias 
which remains constant with time \cite{Desjacques:2010re,Desjacques:2010mo}. If true, 
this would have important consequences for our description of redshift space statistics 
such as the power spectrum \cite{Kaiser:1987cl}. However, there is thus far no evidence 
for such an effect from $N$-body simulations.
Furthermore, the coupled-fluids approximation for the coevolution of dark matter and halos,
which is widely used to compute the time evolution of bias \cite{Fry:1996th, Tegmark:1998wm, Hui:2007ev}, 
predicts that any initial velocity bias rapidly decays to zero \cite{Elia:2010en}. 
This stands in conflict with predictions from the peak approach.

Here, we will show that these two seemingly contradictory results can be reconciled
if one recognizes that the gravitational force acting on biased tracers of the large scale 
structure is itself biased. Consequently, when the purpose is to compute correlators 
like power spectra etc., one should interpret the two-fluids approximation as describing 
effective/mean-field quantities and, therefore, momentum conservation must be modified
accordingly. We will explain how the Euler equation should be modified, and demonstrate that
the new two-fluids approximation predicts the (time) constancy of any {\it statistical} 
velocity bias present in the initial conditions. We will then explore the scale-dependence of
Lagrangian halo bias in simulations, demonstrate the existence of an initial halo velocity 
bias in good agreement with peak theory and present 
numerical evidence that it persists until virialization.

\section{Evolution in the Peak Model}

Virialized halos of mass $M$ can be traced back to the initial conditions to define 
proto-halos, i.e., the Lagrangian patches that will collapse into halos.
In the peak approach, these are associated with peaks of the initial density field (assumed 
Gaussian throughout this letter) smoothed on the halo Lagrangian scale $R\propto (M/\bar{\rho})^{1/3}$. 
As shown in \cite{Desjacques:2008ba,Desjacques:2010re}, the large-scale limit $r\gg 1$ of the peak 
2-point correlation and mean pairwise velocity can be thought of as arising from the effective (or 
mean-field) relations 
$\delta_\text{pk}(\vec x) = b_{10} \delta_\text{m}(\vec x) - b_{01}\nabla^2\delta_\text{m}(\vec x)$
and
$\vec v_\text{pk}(\vec x) = \vec v_\text{m}(\vec x) - R_v^2 \vec\nabla\delta_\text{m}(\vec x)$.
Here,  $b_{10}$ and $b_{01}$ are Lagrangian bias factors whereas $R_v$ is the characteristic scale
of the peak velocity bias.
Hence, we expect the Fourier modes $\delta_\text{h}(\vec k)$ and $\vec v_\text{h}(\vec k)$ of the 
proto-halo overabundance and velocity, respectively, to scale as
\begin{align}
\delta_\text{h}(\vec k)/W_R(k) &= c_1(k) \delta_\text{m}(\vec k) = \left(b_{10}+b_{01} k^2\right)
\delta_\text{m}(\vec k)\, , \\
\vec v_\text{h}(\vec k)/W_R(k) &= b_v(k)  \vec v_\text{m}(\vec k) = \left(1-R_v^2 k^2\right)
\vec v_\text{m}(\vec k) \, ,
\end{align}
at leading order. The peak selection function $W_R(k)$ asymptotes to unity for $k \to 0$ such that, 
on large scales both halo and matter velocity agree.
One usually employs a Gaussian to guarantee convergence of the peak model calculation.
The time evolution of $c_1(k)$ and $b_v(k)$ can be worked out exactly within the Zel'dovich approximation. 
Assuming that the motion of the proto-halo center-of-mass coincides with that of the peak position, 
Ref.~\cite{Desjacques:2010mo} found schematically
\begin{align}
c_{1,\text{pk}}^\text{E}(k,a) &= b_v(k) + D_+^{-1}(a) c_1(k) \label{eq:pklinc1} \\
b_{v,\text{pk}}^\text{E}(k,a) &= b_v(k) \label{eq:pklinbv} \;,
\end{align}
where the superscript E labels evolved quantities and $D_+(a)$ is the linear growth rate normalized 
to unity at the collapse redshift. The linear Lagrangian bias $c_1(k)$ decays to the linear velocity bias, 
which remains constant throughout time.


This prediction stands in sharp contrast with that of the coupled fluid approximation to halos and dark matter. 
In this approach, halos and matter are modelled as pressureless fluids co-evolving in the potential determined 
by the matter distribution solely. Upon linearizing the continuity and momentum conservation equations for 
the Fourier modes of the halo and dark matter density fields, $\delta_\text{h}(\vec k)$ and 
$\delta_\text{m}(\vec k)$, and velocity divergences, $\theta_\text{h}(\vec k)$ and $\theta_\text{m}(\vec k)$, 
the time evolution of any initial $c_1(k)$ and $b_v(k)$ is 
\begin{align}
c_{1,\text{fluid}}^\text{E}(k,a) &= 1 + D_+^{-1}(a) \bigl(c_1(k)+2b_v(k)-3\bigr)\nonumber\\
& +2 D_+^{-3/2}(a) \bigl(1-b_v(k)\bigr) \dots \\
b_{v,\text{fluid}}^\text{E}(k,a) &= 1 + D_+^{-3/2}(a)\bigl(b_v(k)-1\bigr) \;.\label{eq:linvelbiasevol}
\end{align}
In the absence of an initial velocity bias, $b_v^\text{E}=b_v=1$ and we recover the usual decay of the linear 
Lagrangian bias, $c_{1,\text{fluid}}^\text{E}=1+D_+^{-1} (c_1-1)$ \cite{Fry:1996th}. In general, the above equations predict 
the rapid decay of any non-vanishing initial velocity bias.


Can we reconcile these two apparently contradictory results? Firstly, we should bear in mind that this fluid
approximation aims at predicting correlators of halos and dark matter. This is the reason why we are allowed
to relate $\delta_\text{h}$ with $\delta_\text{m}$ through some (local or nonlocal) bias expansion. In other
words, $\delta_\text{h}$ and $\theta_\text{h}$ should in fact be thought of as being {\it effective} or 
{\it mean-field} quantities given a realization of $\delta_\text{m}$, in analogy with the interpretation of 
the peak bias expansion described in \cite{Desjacques:2013eb}.
Consequently, the Euler equation for halos describes the conservation of momentum of the halo {\it mean-field}
density field $\delta_\text{h}(\delta_\text{m})$. The halo momentum conservation equation will be the same 
as that of the dark matter only if the gravitational force acting on halos is {\it statistically unbiased} 
relative to the force acting on dark matter particles. In general, we argue that the Euler equation for halos
should be changed into
\begin{equation}
\frac{\partial\theta_\text{h}(\vec k)}{\partial\eta}+{\cal H}\theta_\text{h}(\vec k)
+\frac{3}{2} b_a(\vec k){\cal H}^2\Omega_m \delta_\text{m}(\vec k) = \text{m.c.}
\label{eq:modeuler}
\end{equation}
Here, $\eta$ is the conformal time, ``m.c.'' designates second-order mode-coupling terms and we have omitted 
the dependence of $\theta$ and $\delta$ on $\eta$. $b_a(\vec k)$ is the 
linear gravitational force or acceleration bias. It is important to note that our modification to the Euler 
equation does not contradict the equivalence principle since it only makes sense statistically: 
on an object-by-object basis, the biased tracers still satisfy the usual momentum conservation.

Since, in the linear regime, the acceleration is parallel to the initial velocity and since gravity mode-coupling 
cannot induce any linear contribution (by definition), $b_a(\vec k)$ must be equal to the linear, initial 
velocity bias $b_v(k)$. Solving the fluid equations
for halos and dark matter with this new halo momentum conservation equation, we find that $c_1^\text{E}(k,a)$
and $b_v^\text{E}(k,a)$ evolve in accordance with the peak predictions Eqs.~\eqref{eq:pklinc1} -- 
\eqref{eq:pklinbv}. To convince ourselves that this must be true, consider the set of points with zero initial 
velocities (e.g. \cite{Percival:2008ga}) as the discrete, biased tracers of the mass density 
field. We thus expect $b_v(\vec k)\equiv 0$ and the linear Eulerian bias $c_1^\text{E}(k,a)$ to decay to zero. 
This is indeed what happens since, in the linear regime, the mass 2-point correlation grows with $D_+^2(a)$ 
while the 2-point correlation of these biased tracers remains constant. 
We note that the mode-coupling terms may also be biased, but we leave this possibility to future studies.
\begin{figure}[t]
\includegraphics[width=0.49\textwidth]{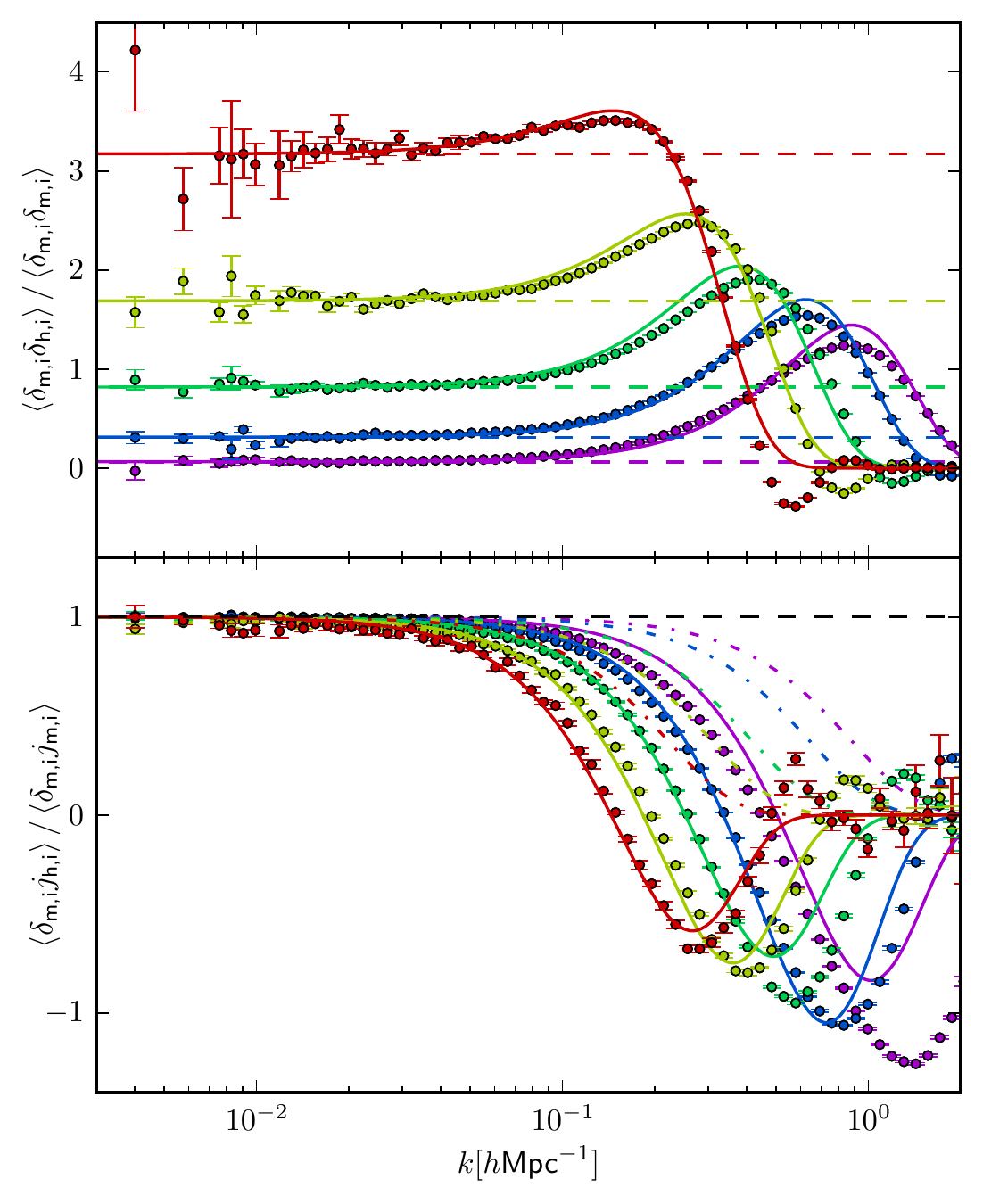}
\vspace{-0.8cm}
\caption{Initial peak density bias $c_1(k)$ (upper panel) and velocity bias $b_v(k)$ (lower panel) at $z_i=99$. 
Having fit $b_{10}$, $b_{01}$ and $R$ from the density bias, we find strong evidence for a non-zero 
$R_v$ in the velocity bias. 
To highlight this detection, we overplot the damping introduced by the 
peak smoothing $R$ alone in the lower panel as dash-dotted lines. The horizontal dashed lines show the 
scale-independent local bias, while the solid lines show the peak model fits. Halo mass is in the range
$8\times 10^{12} - 6\times 10^{14}\hMs$ and increases from bottom to top (top to bottom) in the upper 
(lower) panel.}
\label{fig:bias_init}
\vspace{-0.5cm}
\end{figure}
\section{Initial Velocity Bias in Simulations}
So far, there has not been any conclusive evidence for a statistical velocity bias in the distribution of 
virialized dark matter halos, except for a couple of tentative measurements from \cite{Percival:2008ga,Elia:2011ds}.
These suffered from the fact that simulations are sampling the cosmic density field with a 
finite number of discrete tracer particles. Hence, it is difficult to define a velocity field throughout 
the whole simulation volume, especially for the rare, massive dark matter halos. 
Here, we work instead with the number-weighted (for halos) and density-weighted (for dark matter) velocity
fields \cite{Okumura:2012di}. 
These ``momentum'' fields
$\vec j_\text{h}=(1+\delta_\text{h})\vec v_\text{h}$ and 
$\vec j_\text{m}=(1+\delta_\text{m})\vec v_\text{m}$ 
are well defined everywhere.
To extract statistical information about them, we will measure
the density-momentum correlators $\langle\delta_\text{m} j_\text{m,h}^z\rangle$, where $j_\text{m,h}^z$ is 
the matter or halo momentum projected along the $z$-axis. Since it is a cross-correlation with 
$\delta_\text{m}$, it does not suffer from shot-noise, like the halo-matter cross power spectrum 
$\langle\delta_\text{m}\delta_\text{h}\rangle$. 
We begin by assessing whether both Lagrangian $c_1(k)$ and $b_v(k)$ have $k^2$-dependencies in agreement with 
that predicted by peak theory. In this regard, we consider a suite of 16 collisionless dark matter simulations. 
The initial conditions for the $1024^3$ particles in the $V=1500^3 \hMpcc$ box were set up at 
$z=99$ using 2{\small LPT} \cite{Crocce:2006tr} and subsequently evolved using {\small GADGET}2
\cite{Springel:2005mi}. 
Halos are identified at $z=0$ with a {\small FoF} halo finder of linking length $0.2$ and their 
constituent particles were traced back to the initial conditions to define the Lagrangian 
halo distribution (or proto-halos). We repeat this procedure for a set of intermediate time steps. 
This provides us with the non-linear time evolution of a strictly conserved set of tracers of 
large scale structure.
The mean mass of the five bins is $7.8\tim{12},2.3\tim{13},6.9\tim{13},2.0\tim{14}$ and $5.7\tim{14}\hMs$.

We then measure the ratio of the initial halo-matter and matter-matter density-density and density-momentum 
power spectra. In the peak model, at the lowest order, these quantities are given by
\begin{align}
\frac{\la\delta_\text{m,i}(\vec k)\delta_\text{h,i}(-\vec k)\ra}
{\la\delta_\text{m,i}(\vec k)\delta_\text{m,i}(-\vec k)\ra}
&= \left(b_{10}+b_{01} k^2\right) W_R(k)\\
\frac{\la\delta_\text{m,i}(\vec k) j_\text{h,i}^z(-\vec k)\ra}
{\la\delta_\text{m,i}(\vec k) j_\text{m,i}^z(-\vec k)\ra} &= 
\left(1-R_v^2 k^2\right) W_R(k) \;.
\end{align}
In a first step we fit the linear density bias from the density correlator on large scales, then we jointly fit 
for the scale of the Gaussian peak selection function $R$ and the $k^2$ bias term $b_{01}$ in the same statistic. 
Subsequently we use the filter with the same scale to fit the scale-dependence of the velocity bias and find 
strong evidence for a non-zero initial $R_v$. In App.~\ref{app:velwin} we reconsider the case where the scale dependence of the velocity correlator is entirely due to the window function and find a poor performance for the density statistic, thus strengthening the case for the scenario presented here.
Fig.~\ref{fig:bias_init} shows that this parametrization is able to reproduce the scale-dependence of the proto-halo 
density and velocity bias reasonably well. 
The choice of a Gaussian for the peak selection function is motivated by the sole requirement that the spectral 
moments of the Gaussian field should be convergent. 
This would not be the case for a top-hat window, but generalized window functions might provide a better fit and 
still yield convergent moments. In Appendix \ref{app:massdep} we show that the fitted peak density and velocity bias parameters are in agreement with theoretical predictions once a model for the mass--collapse threshold relationship has been employed. We also show that the coefficients of the $k^2$ corrections are fairly insensitive to the choice of the collapse threshold and thus robust to choices that go beyond the peak constraint. For this study we decide to employ the fitted parameters that describe the initial conditions and to compare the resulting predictions to the evolved halo statistics in the simulations.

\section{Time Dependence of Velocity Bias}
Next, we turn to the time-dependence of the halo velocity bias. To this purpose, we consider the time evolution of 
the linear density-density and density-momentum correlators. 
These linear correlators are obtained by cross-correlating the evolved halo positions and momenta with the linear 
Gaussian matter density field $\delta_\text{m}^{(1)}=D_+(z)D^{-1}_+(z_\text{i})\delta_\text{m,i}$. 
Considering cross-correlations with the non-linear matter field would contaminate the statistics with the poorly 
understood late time matter distribution and, thus, undermine a clear isolation of the scale-dependencies induced by
the peak constraint.
We also refrain from using halo-halo correlators, since these are likely plagued by non-trivial shot noise effects 
\cite{Hamaus:2010mi}.

We assume that peaks move according to their initial velocity as in Zel'dovich approximation. 
We calculate the resulting correlators by writing the evolved peak positions as 
$1+\delta_\text{h}(\vec x)=\bar n_\text{h}^{-1}\sum_\text{h} \ddir(\vec x-\vec x_\text{h})=\bar n_\text{h}^{-1} 
\int \derd^3 q \ddir(\vec x-\vec q-\vec \Psi(\vec q))\sum_\text{pk}\ddir(\vec q-\vec q_\text{pk})$ following 
the steps laid out in \cite{Desjacques:2010mo}, where $\vec \Psi(\vec q)$ is the displacement field at Lagrangian 
position $q$. 
Since the initial matter fluctuations are Gaussian, we only select the linear terms in the bias relation. 
We finally obtain
\begin{align}
\bigl\langle\delta_\text{m}^{(1)}(\vec k) \delta_\text{h}(-\vec k)\bigr\rangle&=D^2_+c_1^\text{E}(k,a)G_\text{pk}(k)P(k)W_R(k)\, ,
\label{eq:densevol}
\\
\bigl\langle\delta_\text{m}^{(1)}(\vec k)  j_\text{h}^z(-\vec k)\bigr\rangle &=
\Bigl(b_{v,\text{pk}}(k)-D^2_+\sigma_\text{d,pk}^2\, c_{1,\text{pk}}^\text{E}(k,a)\,k^2\Bigr) \label{eq:momevol} \\
&\times\mathcal{H}f_+ D^2_+\left(i\frac{\vec k\cdot\zvh}{k^2}\right) G_\text{pk}(k)P(k)W_R(k) \nonumber \, ,
\end{align}
where $c_{1,\text{pk}}^\text{E}(k,a)$ and $b_{v,\text{pk}}(k)$ are defined in Eqs.~\eqref{eq:pklinc1} and \eqref{eq:pklinbv},  $G_\text{pk}(k)=e^{-\frac{1}{2}\sigma_\text{d,pk}^2k^2D_+^2(a)}$ is the peak propagator 
and $\sigma_\text{d,pk}^2$ is the peak displacement dispersion (extrapolated to the collapse epoch), given by 
$\sigma_\text{d,pk}^2=\sigma_{-1}^2-\sigma_0^4/\sigma_1^2$ with $\sigma_i^2=1/3\int d^3k/(2\pi)^3k^{2i}P(k)W_R^2(k)$. 
It is reduced relative to the linear matter displacement dispersion because i) $\sigma_{-1}$ is smaller for halos 
than for matter due to the finite smoothing scale R (which is zero for the matter), and ii) the dark matter preferentially 
flows onto the peaks, so statistically 
the peaks are more at rest than the dark matter, and the term $-\sigma_0^4/\sigma_1^2$ accounts for that.
The analogy with the Eulerian coevolution model in Eqs.~\eqref{eq:pklinc1} -- 
\eqref{eq:pklinbv} can only be seen in the low-$k$ limit due to the resummation of the displacement dispersions.  
In principle non-local third order bias corrections \cite{Biagetti:2014no,Saito:2014un} contribute to the density correlator at late times. 
As we will discuss shortly, these loop corrections are likely suppressed by the explicit smoothing scale in the peak model.
\begin{figure}[t]
\includegraphics[width=0.49\textwidth]{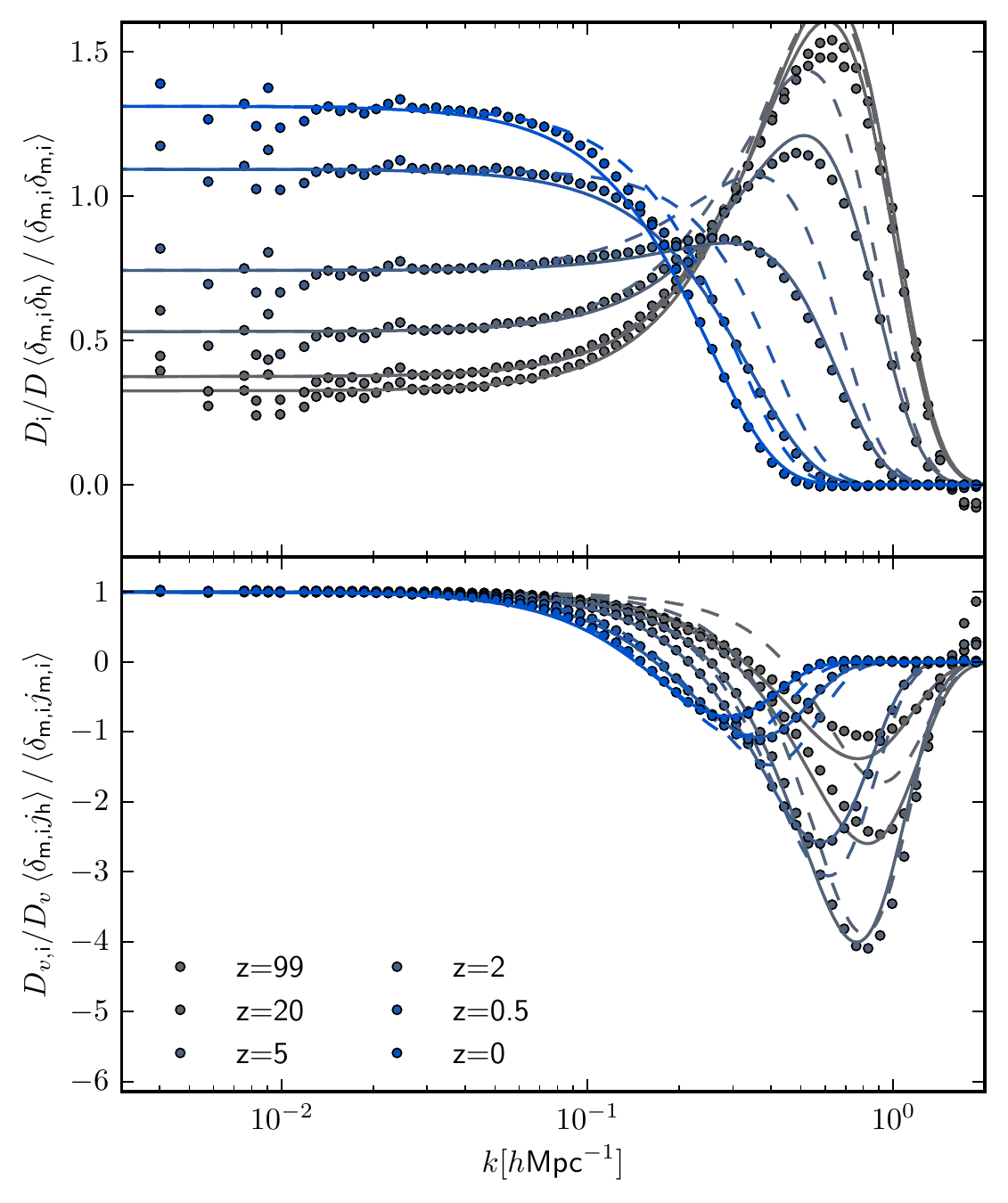}
\vspace{-0.8cm}
\caption{Evolution of the linear density and velocity bias for bin II ($M=2.3\tim{13}\hMs$). $D_v$ is 
the velocity growth factor $D_v=D_+f_+\mathcal{H}$. The lines show our evolution model to linear order, 
Eqs.~\eqref{eq:densevol}--\eqref{eq:momevol}. The dashed lines assume that the initial velocity bias decays
according to the Eulerian co-evolution model, Eq.~\eqref{eq:linvelbiasevol}, and are undoubtedly in 
tension with the simulation data for both statistics even at redshift $z=20$. At low redshifts the damping is dominated by 
the propagator for intermediate wavenumbers, such that this tension becomes less significant for the momentum statistic.}
\label{fig:velbias_evol}
\vspace{-0.3cm}
\end{figure}

\begin{figure*}[t]
\includegraphics[height=10cm]{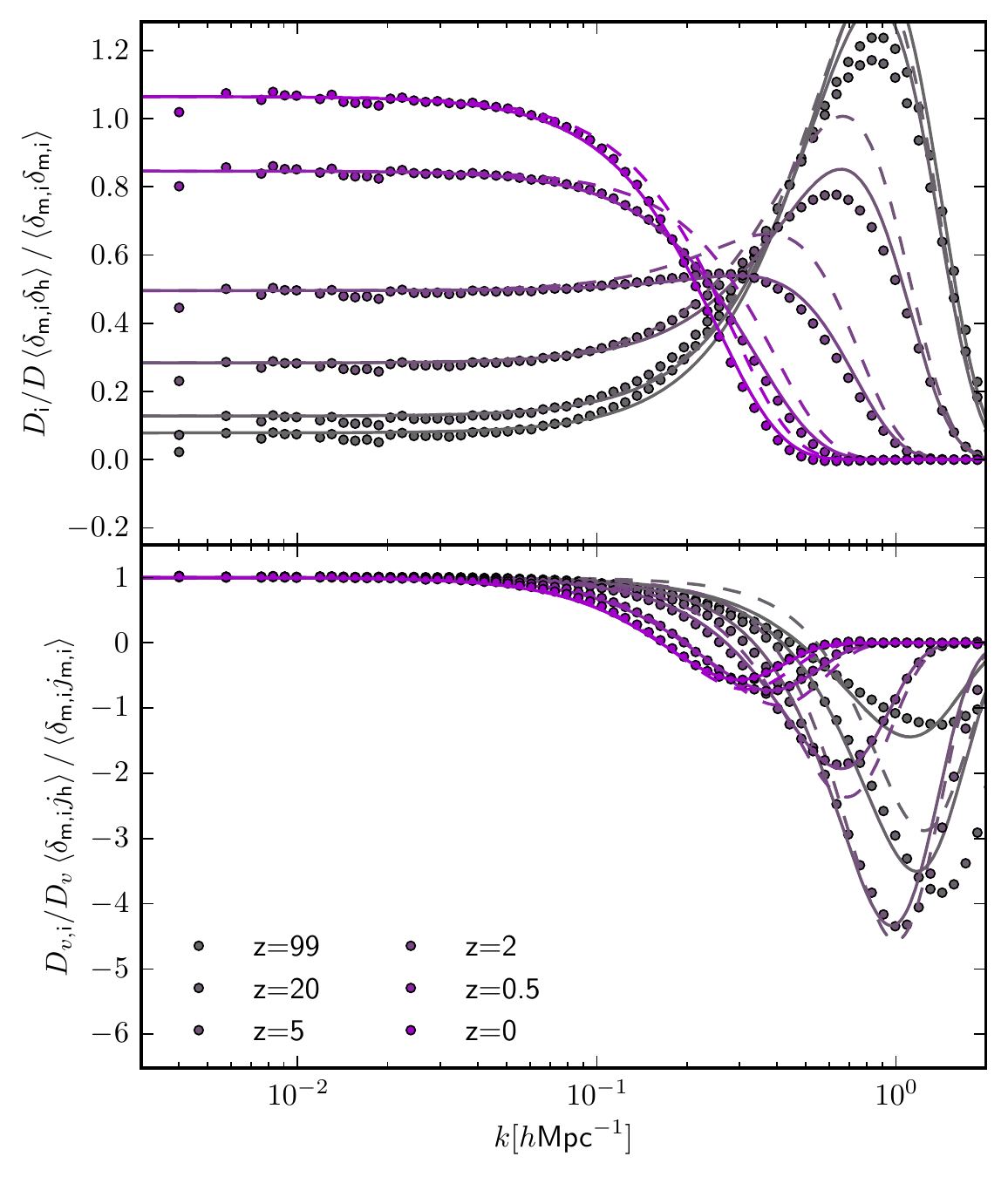}
\includegraphics[height=10cm]{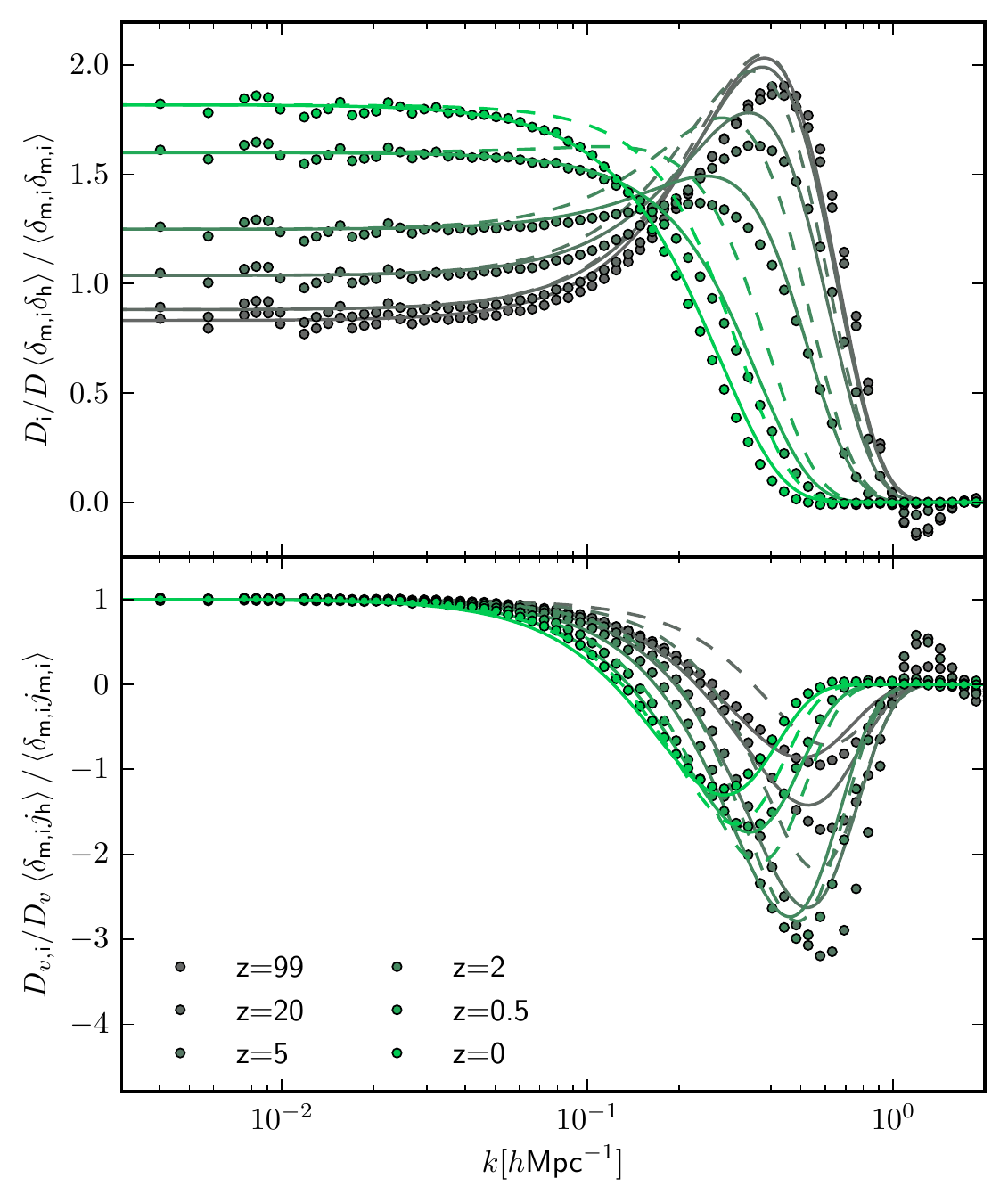}
\includegraphics[height=10cm]{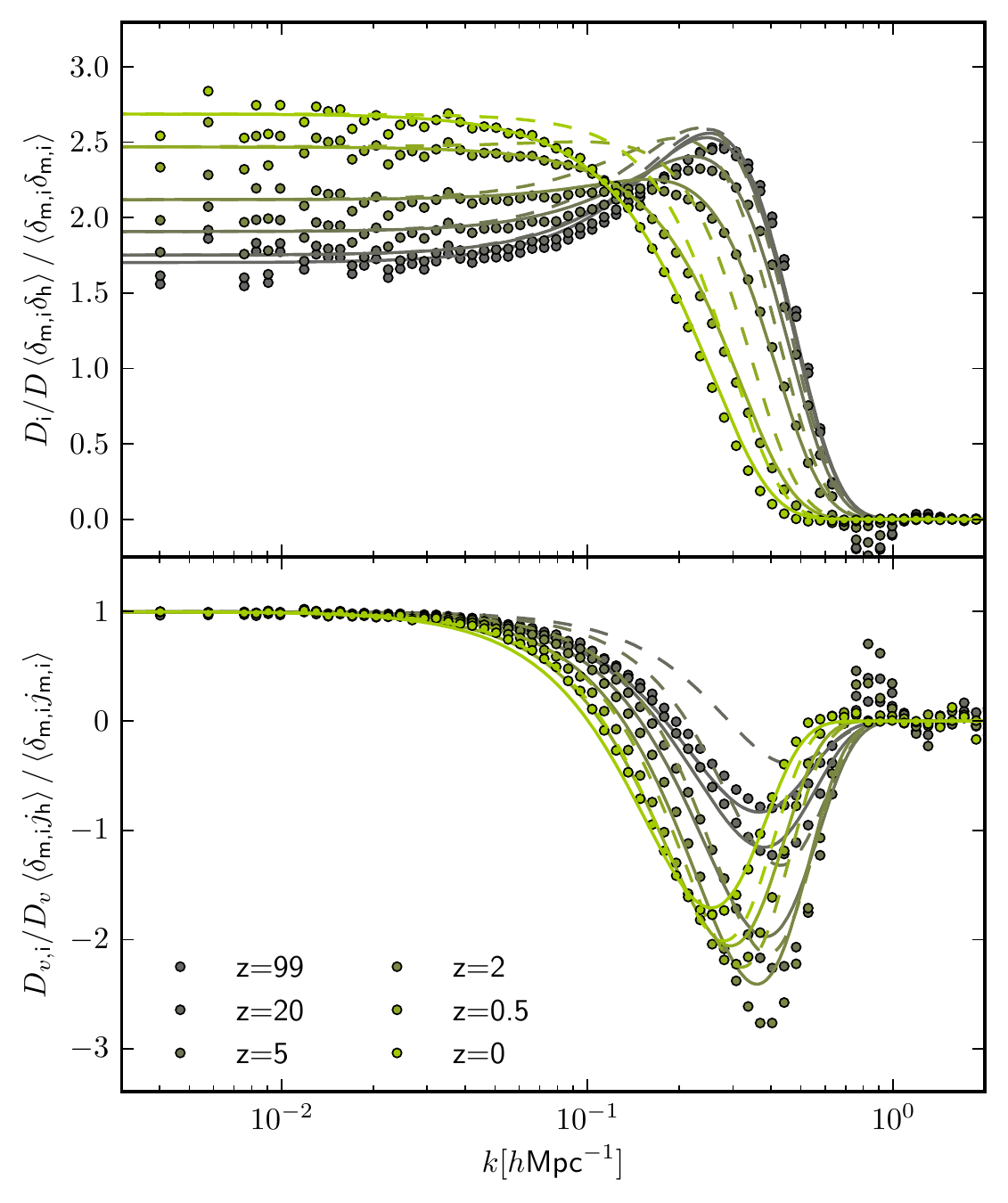}
\includegraphics[height=10cm]{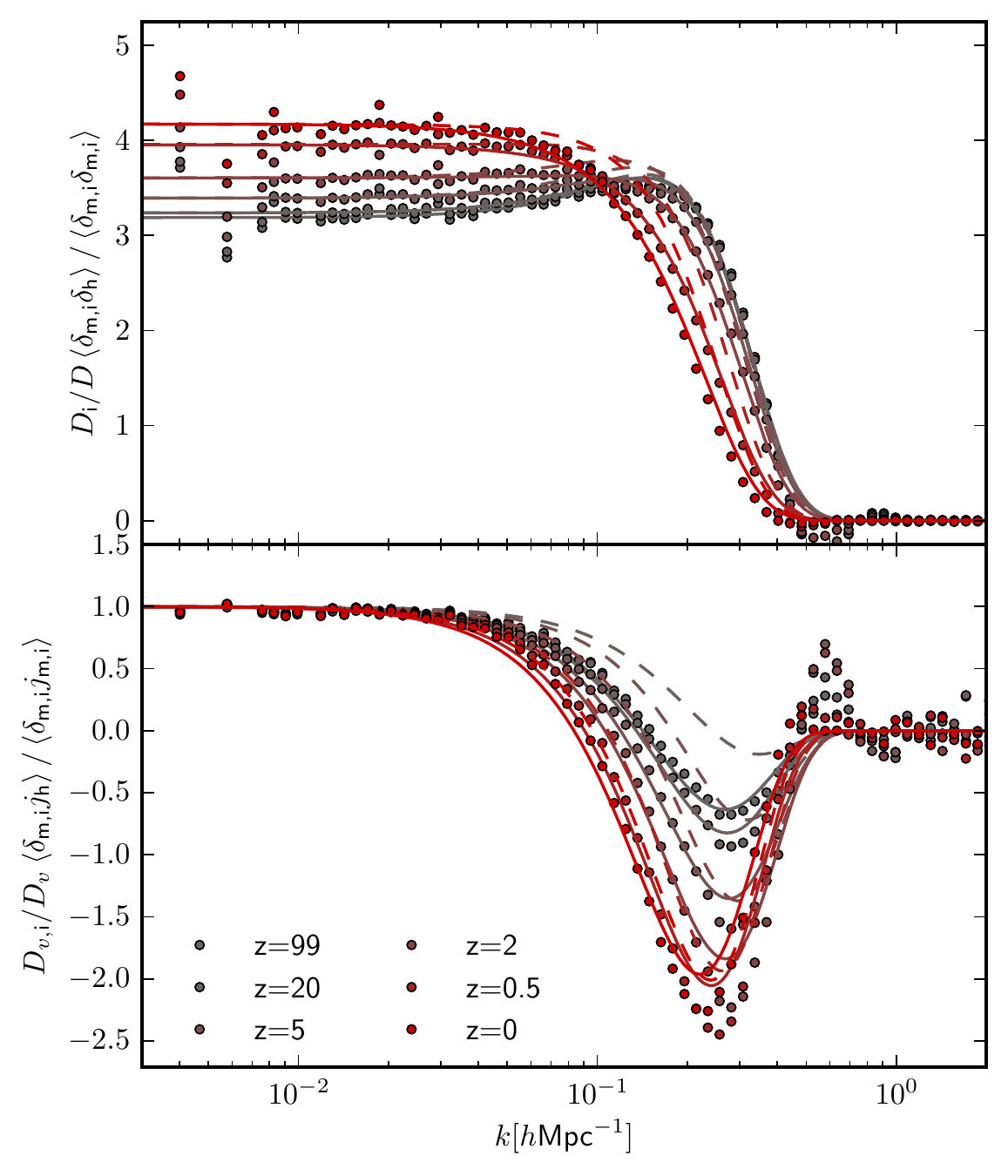}
\vspace{-0.2cm}
\caption{Same as Fig.~\ref{fig:velbias_evol}, just for mass bins I,III,IV and V from top left to bottom right.}
\label{fig:velbias_evol_all}
\vspace{-0.3cm}
\end{figure*}

The evolution of the density and velocity bias in our simulations for mass bin II is shown in 
Fig.~\ref{fig:velbias_evol} and in Fig.~\ref{fig:velbias_evol_all} for the other bins. We divide the cross-correlators of halo density/momentum with the linear matter density by those of the linear matter density/momentum with the linear matter density. On large scales, the density bias in the upper panels evolves from the Lagrangian to the Eulerian bias, i.e. it changes by unity. On smaller scales, most of the initial $k^2$ enhancement in the density bias is washed out by the linear evolution of the peak density bias and the additional damping due to the peak displacement dispersion in the propagator. We show the full prediction of Eq.~\eqref{eq:densevol} (solid lines) and the case in which the velocity bias part of $c_1^\text{(E)}$ decays as $D_+^{-3/2}$ as predicted by coevolution (dashed lines). At all redshifts, the model with decaying velocity bias provides a damping that is too small, thus overpredicting the clustering. The improved agreement in the case of conserved velocity bias is due to a partial cancellation of the $k^2$ coefficient $b_{01}-R_v^2$.

The velocity bias in the lower panels of Figs.~\ref{fig:velbias_evol} and \ref{fig:velbias_evol_all} is unity on large scales as required by mass and momentum conservation. On smaller scales, 
with evolving time we see a sharpening of the dip until redshift $z=5$ when it starts being damped by the exponential propagator arising from the displacement dispersion. 
We show the full prediction of the peak model with conserved velocity bias Eq.~\eqref{eq:momevol} as the solid line.
We also overplot the prediction of the simple linear coevolution model Eq.~\eqref{eq:linvelbiasevol} as dashed lines, where we have kept the exponential part of Eq.~\eqref{eq:momevol} that is not part of the linear evolution but clearly necessary for unequal time statistics. The model with decaying velocity bias is clearly less compatible with the non-linear evolution measured in the data than the model with constant velocity bias. This is particularly evident at intermediate times, where the propagator is not dominant yet. Comparing the various mass bins, we see that the magnitude of the initial and final velocity biases increases with increasing mass.

\begin{figure}
\includegraphics[width=0.49\textwidth]{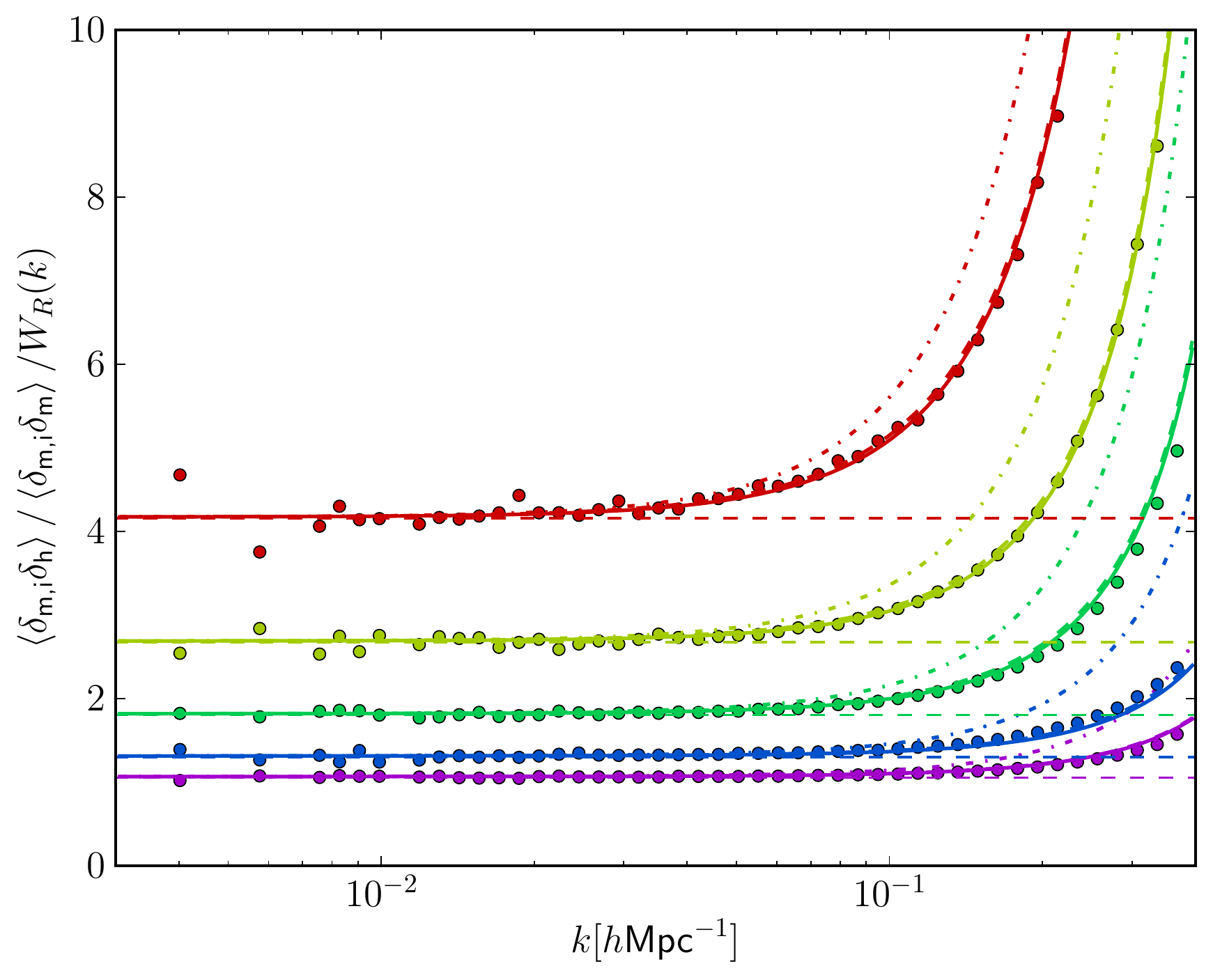}
\vspace{-0.8cm}
\caption{Correlation between the initial Gaussian density field and the final halo density field normalized by the matter propagator. The mass of the haloes is increasing from bottom to top. The horizontal dashed lines show the linear Eulerian bias. The dash-dotted lines show the final peak bias in absence of velocity bias and the solid lines include the velocity bias. The second set of dashed lines that nearly coincides with the solid lines adds the third order non-linear bias \cite{Saito:2014un}, which leaves the results basically unchanged due to the explicit Lagrangian smoothing scale in the loop integrals.}
\vspace{-0.6cm}
\label{fig:finprob}
\end{figure}

In Fig.~\ref{fig:finprob} we show the cross correlation between the final halo density and the linearly 
evolved initial Gaussian matter density field. This quantity asymptotes to the Eulerian bias on large scales. 
On smaller scales, the data show a distinct upturn that is very well described by the peak model in presence 
of undamped velocity bias. The inclusion of non-local third order bias  does not deteriorate this agreement. 
The third order non-local bias would contribute through a loop correction, i.e. an integral over the power 
spectrum weighted by a kernel $K_{3,\text{nl}}$
\be
P_{b_3,\text{nl}}(k)\propto P(k)W_R(k) \int \frac{\derd^3 q}{(2\pi)^3}P(q)K_{3,\text{nl}}(\vec q,-\vec q,\vec k)W_R^2(q)\; .
\ee
In contrast to standard implementations \cite{Saito:2014un,Biagetti:2014no}, we explicitly account for the 
smoothing scale in the relation between halo overdensity and matter, which significantly suppresses the loop 
corrections. In practice, we have adopted the third order non-local bias amplitude as predicted by Eulerian 
coevolution $b_{3,\text{nl}}\propto b_{10}$, and set the smoothing scale to the Lagrangian scale.

\section{Conclusions}
We have focused on modeling correlators 
with the linear, Gaussian density field in order to isolate the effects of the scale-dependent, linear peak
bias. These statistics are, admittedly, not those that will be directly measured in observations. However, 
they allow us to understand key physical properties of the joint evolution of halos and matter under gravity and, 
therefore, they will help predicting the true observables more accurately and reliably. 
Our result also demonstrates that, at the level of linear bias, the Zel'dovich evolved peaks provide a very good
description of the time evolution of the halo-matter correlators. While the interior structure of a dark matter 
halo is highly non-linear, the effective dynamical evolution of its center of mass is more amenable to perturbative 
treatments than the dark matter itself.
Finally, we insist on the interpretation of $\delta_\text{h}$, $\theta_\text{h}$ in the halo-matter 
fluid approximation as mean-field/effective quantities (and not counts-in-cells), as long as the purpose 
is to compute ensemble averages.

Our measurement of a halo velocity bias constant throughout time may have important 
consequences for the modeling of halo clustering in redshift space, see e.g.~\cite{Desjacques:2010re}.
However, further developments are required until peak theory can provide an accurate template for the halo-halo 
correlation function. 
Since the full, non-perturbative initial peak-peak correlation function correctly accounts for exclusion effects 
\cite{Baldauf:2013ha}, it should lead to a comprehensive description of all the scale-dependent bias effects. 
On a final note, the effective field theory descriptions of halo statistics \cite{McDonald:2009cl,Assassi:2014re} 
may lead to a functional form similar to that predicted 
by the peak model, but the model presented in this work complements it by a dynamical perspective and provides
physical arguments for the values of the operators.

\begin{acknowledgements}
The authors would like to thank Roman Scoccimarro, Ravi Sheth and Zvonimir Vlah for useful discussions.
T.B. gratefully acknowledges support from the Institute for Advanced Study through the W. M. Keck Foundation Fund.
V.D. acknowledges support by the Swiss National Science Foundation.
U.S. is supported in part by the NASA ATP grant NNX12AG71G.
\end{acknowledgements}

\begin{figure*}[t]
\includegraphics[width=0.49\textwidth]{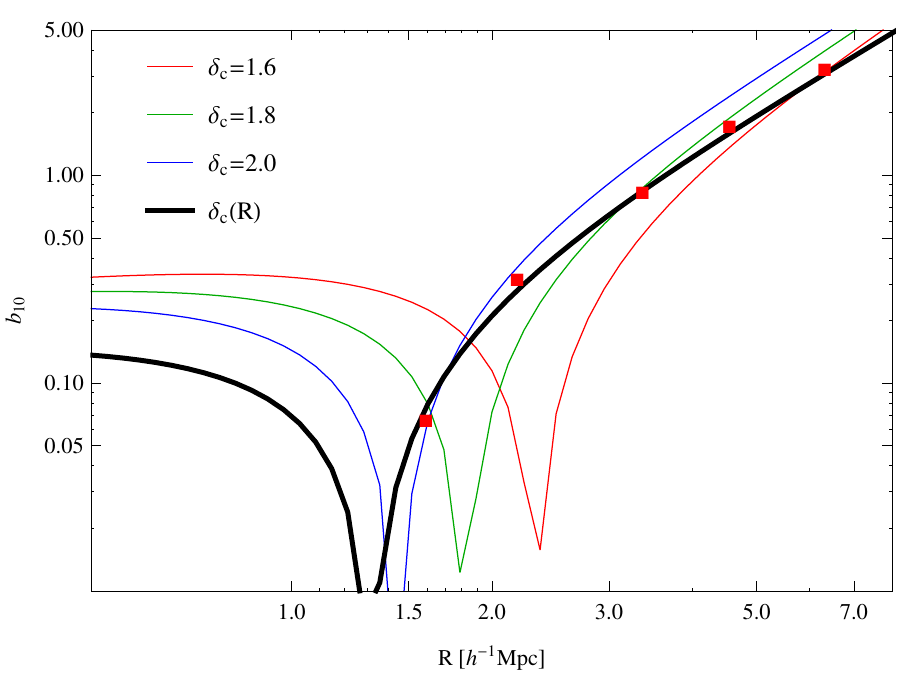}
\includegraphics[width=0.49\textwidth]{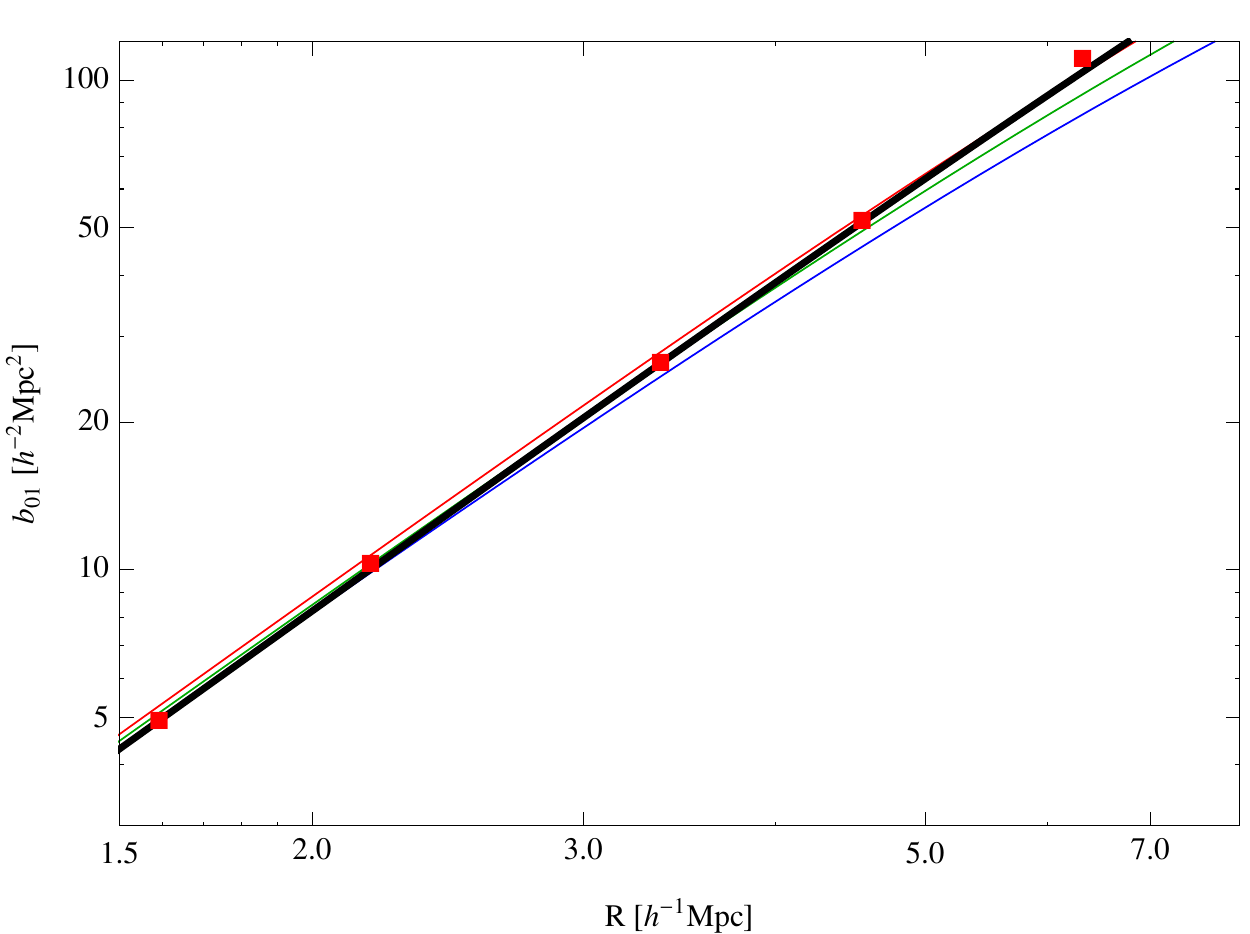}
\includegraphics[width=0.49\textwidth]{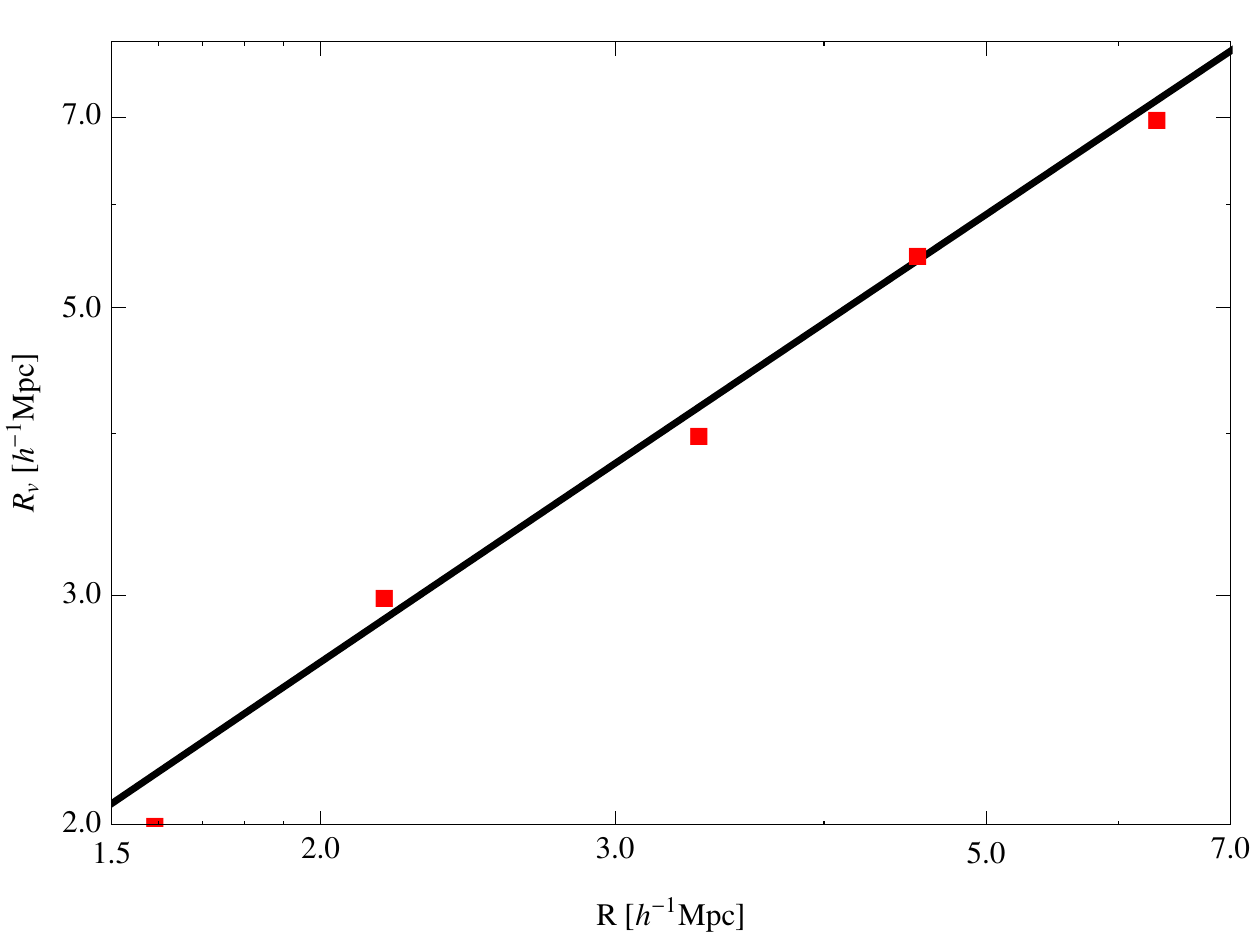}
\caption{Bias parameters derived from the peak model. \emph{Top left panel: }$k^0$ density bias. \emph{Top right panel: }$k^2$ density bias \emph{Bottom panel: }$k^2$ velocity bias. The points show the measurements in the initial conditions, the colored lines show the predictions for three fixed collapse thresholds and the thick black line shows the predictions based on our model for the mass dependence of the collapse threshold Eq.~\eqref{eq:collapsethreshold}.
Note that only the $k^0$ density bias $b_{10}$ is sensitive to the mass dependence of the collapse threshold.}
\label{fig:biasparameters}
\end{figure*}
\begin{figure*}[t]
\includegraphics[width=0.49\textwidth]{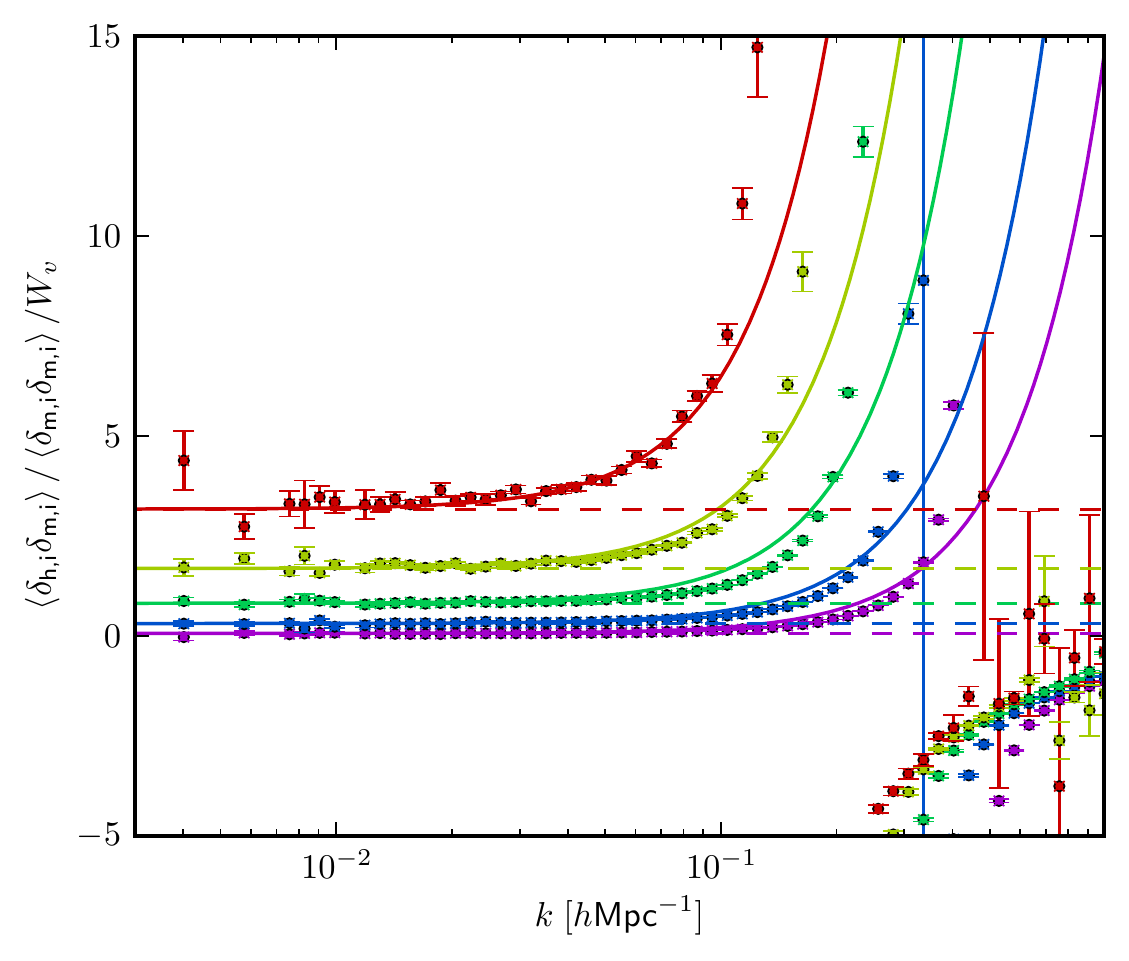}
\includegraphics[width=0.49\textwidth]{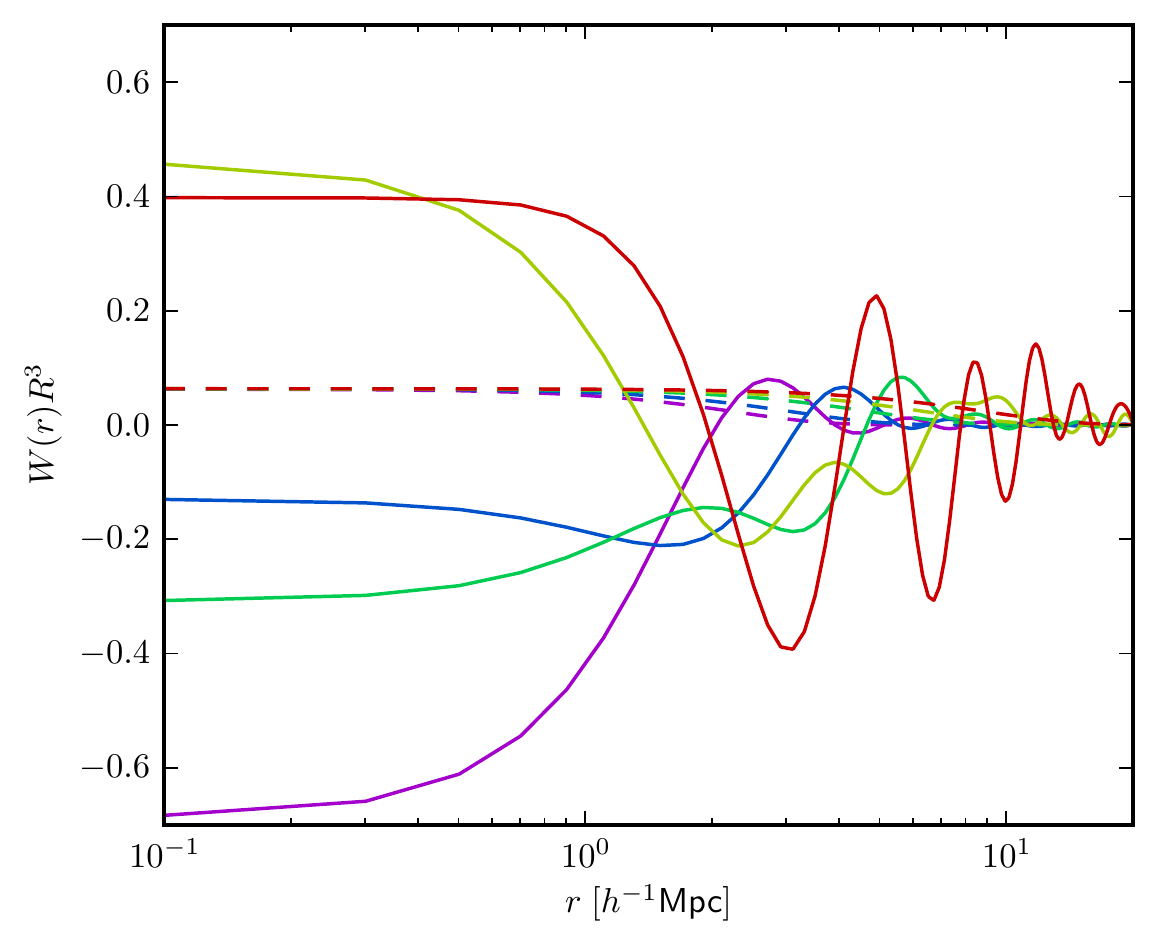}
\caption{\emph{Left panel: }Ratio of the initial density bias and the velocity window. Clearly this functional form can only be fitted by a constant and a $k^2$ term up to the zero crossing of the velocity window, which happens at $k=0.2\ihMpc$, where the model fails. This range is significantly smaller than the range over which the Gaussian window works. \emph{Right panel: }Real space version of the fiducial Gaussian window (dashed) and the velocity window (solid). The latter has a strongly oscillatory behaviour, which is clearly undesirable for a meaningful window function.}
\label{fig:velwindow}
\end{figure*}

\appendix
\section{Peak Bias Parameters}\label{app:massdep}
Predicting the peak bias parameters from the model requires one to choose a relation between halo mass and collapse threshold. While the spherical collapse model suggests a unique collapse threshold of $\delta_\text{c}=1.686$ regardless of the halo mass, measurements in simulations by \cite{Robertson:2009co} indicate that the top hat, linear overdensity around proto-haloes is increasing with decreasing mass. The value of $1.686$ is attained only for very high mass objects. The top hat window can not be used for calculations in the peak model since it does not yield convergent higher derivative moments $\sigma_i, i>0$. A Gaussian window or a mixture of top hat and Gaussian filters can be employed to bypass this problem.
A measurement of the initial overdensity smoothed with a Gaussian window was performed by \cite{Elia:2011ds} for FoF halos. They provided a fitting function, and also found that the collapse threshold increases with decreasing mass. In contrast to  \cite{Elia:2011ds} however, we decided not to tighten the filter scale to the halo mass scale, but rather adjust it so that we reproduce the cutoff in the initial condition density and velocity power spectra. The disagreement between the filter scale and the halo mass likely originates from the fact that the Gaussian filter does not reproduce the proto-halo mass density profile very well. One may want to consider mixed filtering like \cite{Paranjape:2013} or a more general window that interpolates between a Gaussian and a top hat. This is clearly beyond the scope of this paper.\\
We use the following collapse threshold as a function of radius
\be
\delta_\text{c}(R)=2.17 - 0.32\log({R/1 \hMpc}).
\label{eq:collapsethreshold}
\ee
Note that, on rescaling our radii to coincide with the ones predicted by the mass scale, our approximation would be close to the average collapse threshold measured by \cite{Elia:2011ds}.
For the mean peak curvature we rely on \cite{Bardeen:1985th}
\be
\bar u=\gamma \nu + \frac{3 (1 - \gamma^2) + (1.216 - 0.9 \gamma^4) e^{-(\gamma/2) (\gamma \nu/2)^2}}
{\sqrt{3(1-\gamma^2)+0.45 + (\gamma \nu/2)^2} + \gamma \nu/2}
\ee
With these ingredients the peak bias parameters can be calculated according to \cite{Desjacques:2013eb}
\begin{align}
b_{10}=&\frac{1}{\sigma_0}\frac{\nu-\bar u \gamma}{1-\gamma^2}\, ,\\
b_{01}=&\frac{1}{\sigma_2}\frac{\bar u-\nu \gamma}{1-\gamma^2}\, ,\\
R_v^2=&\frac{\sigma_0^2}{\sigma_1^2}.
\end{align}
Note that the velocity bias is independent of the collapse threshold and solely depends on the size of the proto-halo patch. We show the bias parameters in Fig.~\ref{fig:biasparameters} as a function of filter scale. The $k^2$ density bias and the velocity bias are not sensitive to our choice of collapse threshold. We thus consider their amplitude to be fairly model independent. By contrast, the $k^0$ density bias strongly depends on the collapse threshold, which is rather unsurprising since it encodes the response to long wavelength density fluctuations. 

To summarize this section, let us emphasize that the model parameters can be predicted from first principles with reasonable accuracy so long as one adds a few ingredients in addition to the peak constraint. We still consider the evolution predictions based on initial condition measurements in the simulations to be a decisive test of the persistence of velocity bias that is independent of these additional assumptions.


\section{No velocity bias}\label{app:velwin}
In the main text we have fitted the initial density and momentum correlators assuming the functional form of the peak model. In this Appendix, we would like to consider the case, in which the scale dependence of the momentum correlator is entirely due to the window, i.e., that there is no velocity bias beyond the averaging over the halo patch. In this case the momentum correlator defines the ``velocity'' window
\be
W_v(k)\equiv \frac{\la\delta_\text{m,i}(\vec k) j_\text{h,i}^z(-\vec k)\ra}
{\la\delta_\text{m,i}(\vec k) j_\text{m,i}^z(-\vec k)\ra} \;
\ee
In real space, this window has some odd properties, in particular it is not positive definite and oscillatory for most of the mass bins (see right panel of Fig.~\ref{fig:velwindow}). We can now employ this window to describe the protohalo-matter correlator as $(b_{10}+\tilde b_{01} k^2)W_v(k)P(k)$, where the change in the window function induces a change in the $k^2$ part of the bias. The result is shown in the left panel of Fig.~\ref{fig:velwindow}. Dividing by the window, the scale dependence can initially be fitted by a $k^2$ term, but this approach fails once the velocity window crosses zero at $k\approx 0.2 \ihMpc$. This is a considerably smaller range than the range over which the Gaussian window together with the scale dependent density and velocity bias presented in the main text can describe the measurements.

\bibliography{peak_letter}

\end{document}